\documentclass[slac_one]{revtex4}
\usepackage{graphicx}
\usepackage{latexsym}
\usepackage{hyperref}
\usepackage{fancyhdr}
\usepackage{times}
\newsavebox{\savepar}
\newenvironment{boxit}{\begin{lrbox}{\savepar}
\begin{minipage}[b]{4in}}
    {\end{minipage}\end{lrbox}\fbox{\usebox{\savepar}}}

    \newcommand{\eprint}[1]{\mbox{\href{http://arxiv.org/abs/#1}{\tt #1}}}

    \newcommand{\onetev}{1-TeV scale}

    \newcommand{\mev}{\hbox{ MeV}}
    \newcommand{\gev}{\hbox{ GeV}}
    
    \newcommand{\cm}{\hbox{ cm}}

    \newcommand{\cfrac}[2]{\textstyle{\frac{#1}{#2}}}
    
    \newcommand{\mm}{\hbox{ mm}}
    \newcommand{\eqn}[1]{(\ref{#1})}

\preprint{FNAL--CONF--04/nnn--T}

\begin{document}

\title{Nature's Greatest Puzzles}

%

\author{Chris Quigg}
\affiliation{Fermi National Accelerator Laboratory,  P.O. Box 500, 
Batavia, Illinois 60510 USA}

\begin{abstract}
Opening lecture at the 2004 SLAC Summer Institute.

\end{abstract}

\maketitle

\thispagestyle{fancy}


\section{QUESTIONS GREAT AND SMALL}
It is a pleasure to be part of the SLAC Summer Institute again, not simply because it
is one of the great traditions in our field, but because this is a
moment of great promise for particle physics. I look forward to 
exploring many opportunities with you over the course of our two weeks 
together.
My first task in talking about Nature's Greatest Puzzles, the title 
of this year's Summer Institute, is to 
deconstruct the premise a little bit. 

\subsection{The Nature of Scientific Questions \label{subsec:natq}}

        ``About 500 years ago man's curiosity took a special turn toward
        detailed experimentation with matter,'' wrote Viki 
        Weisskopf~\cite{viki}.  ``It was the beginning of
        science as we know it today.  Instead of reaching directly at the
        whole truth, at an explanation for the entire universe, its
        creation and present form, science tried to acquire partial truths
        in small measure, about some definable and reasonably separable
        groups of phenomena.  
    
        ``Science developed only when men began to
        restrain themselves not to ask general questions, such as: What is
        matter made of?  How was the Universe created?  What is the essence
        of life?  They asked limited questions, such as: How does an object
        fall?  How does water flow in a tube?  etc.  Instead of asking
        general questions and receiving limited answers, they asked limited
        questions and found general answers.''
	
An important part of what we might do in these two weeks together is to think 
about how we actually construct science, how we construct 
understanding, and how we present the acts of doing science to other 
people.
Galileo, the icon of the moment when we humans found the courage to
reject authority and learned to interrogate nature by doing
experiments, expressed his approach in this way~\cite{galileo}: 
\begin{quote}
    Io stimo pi\`{u} il trovar un vero, bench\`{e} di cosa leggiera, ch'l
    disputar lungamente delle massime questioni senza conseguir verit\`{a}
    nissuna.\footnote{I attach more value to finding a fact, even about the
    slightest thing, than to lengthy disputations about the Greatest
    Questions that fail to lead to any truth whatever.}
\end{quote}
We have built up science over these past five 
hundred years  not so much by focusing on the majestic questions 
as by thinking about small questions that we have a chance to answer, 
and then trying to weave the answers to those questions together into 
an understanding that will give us insight into the largest questions.

By focusing on
``small things,'' with an eye to their larger implications, Galileo
achieved far more than the philosophers and theologians who surrounded 
him in Florence and Venice, and who, by their  authority,
asserted answers to the ``greatest questions.''  A great shame of the
race of physics professors is that going through Galileo's motions,
\textit{without} an eye to their larger implications, too often
constitutes freshman physics lab. We owe it to our students to explain 
\textit{why} we require them to re\"{e}nact Galileo's investigations, 
how we seek to weave the answers to small questions into broader 
understanding, and what science really is. There is a glorious story 
here, and  we need to convey that glorious story to our 
students and to the public at large. We owe no less to the future 
of our science!

I don't underestimate the value of grand themes as organizing
principles and motivational devices, but I want to emphasize the need
to balance the grandeur and sweep of the Great Questions with our
prospects for answering them.  At every moment, we must decide which
questions to address.  Unimagined progress may flow from small
questions.  Measuring how the conductivity of the atmosphere varies
with altitude, Victor Hess discovered the cosmic
radiation~\cite{SMballoon}---one of the wellsprings of particle physics
and the subject of Great Puzzle No.~9 at this XXXII SLAC Summer
Institute.  Hess did not set out to found particle physics, nor even to
explore the Great Beyond, but merely to pursue a puzzling observation.
So it's entirely possible that by paying close attention to a
\textit{well-chosen small thing,} we may be able to change the world.

I am insisting---with Weisskopf and Galileo and many others---on the
importance of small questions because their role in the making of
science is so poorly understood.  Introducing \textit{Time} Magazine's
top eighteen (not just ten!)  list of America's Best in Science and
Medicine, Michael Lemonick wrote in 2001~\cite{mikel},
\vspace*{-6pt}
\begin{quote}
    ``The questions scientists are tackling now are a lot narrower than
    those that were being asked 100 years ago. \ldots As John Horgan
    pointed out in his controversial 1997 best seller, \textit{The End of
    Science,} we've already made most of the fundamental discoveries: that
    the blueprint for most living things is carried in a molecule called
    DNA; that the universe began with a Big Bang; that atoms are made of
    protons, electrons and neutrons; that evolution proceeds by natural
    selection.''  
\end{quote}
Horgan's assertion that most of the great questions have already been
answered is a relatively puerile form of millennial
madness.  Perhaps this misperception lingers because when we scientists
talk about our work we don't always situate our immediate goals within
a larger picture that would give an image of what we're trying to
learn, what we're trying to understand.  But the notion that science's 
best days are behind us will pass, if it hasn't
already.\footnote{Now that  Mr.~Lemonick has written a biography of
the Wilkinson Microwave Anisotropy Probe~\cite{echos}, I trust that he
has found at least one counterexample!} I'm more troubled by the breezy
claim (``more and more about less and less'') that we scientists today
address narrower questions than our ancestors did a century ago.  This
is preposterously false; it has nothing to do with the way science is
actually done.  Ever since Galileo, what we call science has advanced
precisely by asking, and answering, limited questions, seeking small
facts, and synthesizing an ever-more-comprehensive understanding of
nature.  It is vexing to hear this misconception from a distinguished
science writer.  It is even more vexing because the writer's father was
a legendary Princeton physics professor---and a particle physicist.  We
are failing to communicate that science is, in its essence, 
weaving together the answers to small questions, and we must do better!

Now let us turn for a moment to the list of ``Greatest Puzzles'' that will
command our attention for these two weeks:
    \begin{enumerate}
        \addtolength{\itemsep}{-9pt}\centering
    
        \item  Where and what is dark matter?
    
        \item  How massive are neutrinos?
    
        \item  What are the implications of neutrino mass?
    
        \item  What are the origins of mass?
    
        \item  Why is there a spectrum of fermion masses?
    
        \item  Why is gravity \underline{so} weak?
    
        \item  Is Nature supersymmetric?
    
        \item  Why is the Universe made of matter and not antimatter?
   
        \item  Where do ultrahigh-energy  cosmic rays come from?
    
        \item  Did the Universe inflate at birth?
    \end{enumerate}
To their credit, the
organizers have given you ten ``Greatest Puzzles'' that are not all
Great Questions. Some of them are small questions that might grow, in 
the spirit of Hess's studies of the atmosphere, into great answers.  
I think it's important to recognize
that ``top-ten'' lists\footnote{The essential psychosocial capital 
 that lists generate has been examined by Louis 
Menand~\cite{menand}.} are always subjective in some way: they suit a certain
moment, a certain purpose, a certain institution, a certain prejudice.

It's also true that the list of ``Greatest Puzzles'' changes with time.
To me, one of the most inspiring things about the progress of science
is the way in which questions that were, not so long ago,
``metaphysical''---that couldn't be addressed as scientific 
questions---have become scientific questions.  I give you two that in former times were used 
exclusively to torture graduate students on their
qualifying exams: 
\begin{quote}
    What would happen if the mass of the proton
    or the mass of the electron changed a little bit? \\ What would happen
    if the fine structure constant changed a little bit?  
\end{quote}
When I was on the
receiving end of those questions, I had little patience for them. To 
tell the truth, I really hated them, because the
world wasn't that way, so why think about it?  Now that I've lost some of
the certainty of youth,\footnote{The scaling laws I derived with Jon 
Rosner~\cite{Quigg:1979vr} may be seen as an act of penance for my youthful 
intolerance.} I've come to understand that these were much better
questions than my teachers realized. 

Let's  recast them slightly, as
\begin{quote}
    Why is the proton mass 1836 $\times$ the electron mass? \\
    What accounts for the different strengths of the strong, weak, and
    electromagnetic interactions?
\end{quote}
Not so long ago, these were metaphysical questions beyond the reach of
science:  Masses and coupling strengths were givens.  But now we can
see how the values of masses and coupling strengths might arise; we
recognize these questions as scientific questions.  As we'll recall in
a few paragraphs, we understand where the proton mass comes from.  We
have a framework for inquiring into the origin of the electron mass.
We know, through renormalization group analysis, that coupling
constants evolve with energy; we can make a picture in which the
coupling constants have the low-energy values we measure because they
evolve from a common value at a high energy---the unification scale.
We can imagine how, if the world were a little different, the couplings
would have changed.  So these turn out to be not such annoying
questions---not mere instruments of torture---but questions that we can
 answer scientifically.  Soon, we will be able at least to
sketch plausible storylines, if not to tell the full stories.
Similar progressions from apparently arbitrary givens to answerable
scientific questions appear all over the map of science.

Some questions remain unanswered for so long that we might be tempted 
to forget that they are questions. One that has been much on my mind 
of late is, ``Why are charged-current weak interactions 
left-handed?'' Nearly everyone in this room was born---or at least born 
as a physicist---after the 1957 discovery of parity violation in the weak 
interactions. It's fair to say that, whereas our ancestors were shaken 
by the asymmetry between left-handed and right-handed particles, we 
have grown up with it. [I estimate that I have written down more than 
ten thousand left-handed doublets to this point in my career.] So it would 
not be astonishing if the question had lost its edge for us. But I 
hope you will agree that the distinction between left-handed and 
right-handed particles is one of the most puzzling aspects of the 
natural world. It suggests the following
\begin{center}
    \begin{boxit}
    \textit{Exercise.}  What other profound questions have been with 
    us for so long that they are less prominent in ``top-ten'' lists 
    than they deserve to be?
    \end{boxit}
\end{center}

If new questions come within our reach and long-standing questions slip from our
consciousness, some formerly Great Questions now seem to us the wrong
questions.  A famous example, developed in detail by Lincoln 
Wolfenstein last year~\cite{lincoln}, is Kepler's quest to understand why the
Sun should have exactly six planetary companions in the observed
orbits.  Kepler sought a symmetry principle that would give
order to the universe following the Platonic-Pythagorean tradition.
Perhaps, he thought, the six orbits were determined by the five regular solids of
geometry, or perhaps by musical harmonies.  We now know that the Sun
holds in its thrall more than six planets, not to mention the
asteroids, periodic comets, and planetini, nor all the moons around
Kepler's planets.  But that is not why Kepler's problem seems
ill-conceived to us; we just do not believe that it should have a
simple answer. Neither symmetry principles nor stability criteria make
it inevitable that those six planets should orbit our Sun precisely 
as they do. I think this example holds two lessons for us: First, it 
is very hard to know in advance which aspects of the physical world 
will have simple, beautiful, informative explanations, and which we 
shall have to accept as ``complicated.'' Second, and here Kepler is a 
particularly inspiring example, we may learn very great lessons 
indeed while pursuing challenging questions that---in the end---do not 
have illuminating answers.

Sometimes we answer a Great Question before we recognize it as a
scientific question.  A recent example is, ``What sets the mass of the
proton?''  and its corollary, ``What accounts for the visible mass of
the universe?''  Hard on the heels of the discovery of asymptotic
freedom,\footnote{See the 2004 Nobel Lectures by David
Gross~\cite{dgnobel}, David Politzer~\cite{hdpnobel}, and Frank
Wilczek~\cite{fwnobel}.} Quantum Chromodynamics provided the insight:
the mass of the proton is given mostly by the kinetic energy of three
extremely light quarks and the energy stored up in the gluon field that
confines them in a small space.  Almost before most people realized
that QCD had made the question answerable, we had in our hands the
conceptual answer and an essentially complete \textit{a priori}
calculation~\cite{Chivukula:2004hw} .

\subsection{``The Theory of Everything'' \label{subsec:TOE}}
I do not have a lot of patience for debates about the problem of 
knowledge; for the most part, I would rather do science than talk 
about how to do it. Nevertheless, at this time when we anticipate a 
great flowering of our subject, we should examine our habits and think 
a little bit about how other people do science and how they see us.

Two interesting characters, Bob Laughlin and David Pines~\cite{pandl},
have published a broadside proclaiming the end of reductionism (``the
science of the past''), which they identify with particle physics, and
the triumph of emergent behavior, the study of complex adaptive systems
(``the physics of the next century'').  The idea of emergent behavior,
which they advertise as being rich in its applications to condensed
matter physics in particular, is that there are phenomena in nature, or
regularities, or even very precise laws, that you cannot recognize by
starting with the Lagrangian of the Universe.  These include situations
that arise in many-body problems, but also situations in which a simple
perturbation-theory analysis is not sufficient to see what will happen.

My first response to Laughlin \& Pines is that they have profoundly
misconstrued the way we work.  What is quark confinement in QCD, the
theory of the strong interactions, if not emergent behavior?  You could
do perturbation theory for a very long time and not
discover the phenomenon of confinement.
This notion of emergence is ubiquitous in particle physics.  As QCD
becomes strongly coupled, new phenomena emerge---not only confinement,
but also chiral symmetry breaking and the appearance of Goldstone
bosons---that we wouldn't have anticipated by staring at the
Lagrangian.  [This is, by the way, one of the reasons that we should
force ourselves to pay attention to heavy-ion collisions at high
energies; the very lack of simplicity may push us into realms of QCD where we can't guess
the answers by simple analysis.]  The ``Little Higgs'' approach to
electroweak symmetry breaking~\cite{mp2} is another example of important features
that are not apparent in the Lagrangian in any simple sense.  A
graceful description of the consequences of these phenomena entails new
degrees of freedom and a new effective theory.

Laughlin and Pines advocate the search for ``higher organizing
principles'' (perhaps universal), relatively independent of the
fundamental theory.  I give them credit for emphasizing that many
different underlying theories may lead to identical observational
consequences.  But they turn a blind eye to the idea that in many
important physical settings, the detailed structure and parameters of
the Lagrangian are decisive.  They campaign as well for the synthesis
of principles through experiment, which I also recognize as part of the
way we do particle physics.  I believe that the best practice of
particle physics---of physics in general---embraces both reductionist
and emergentist approaches, in the appropriate settings.

Overall, I am left with the impression that Laughlin \& Pines are
giving a war to which no one should come, because the case for their
revolutionary intellectual movement is founded on misperception and
false choices.\footnote{It is a delicious irony that string theorists,
whose top-down style seems particularly vexing to Laughlin \& Pines and
their allies, may turn out to be---if landscape ideas take hold or spacetime 
is emergent---the
ultimate emergentists!} Perhaps the best way for us to be heard is to
listen more closely, try to understand the approaches we have in
common, and---occasionally---to use their language to describe what we
do.  It is important for us to seek the respect and understanding of
our colleagues who do other physics, in other ways.

One question of scientific style remains: when we understand a
phenomenon as emergent, will that stand as a final verdict, or does
emergence represent a stage in our understanding that will be
supplanted as we gain control over our theories and the methods by
which we elaborate their consequences?  And does one perspective or 
another limit our ability to advance our understanding?

\subsection{Some Other Meta-Questions \label{subsec:metaq}}
I would like to bring these introductory remarks to a close by 
pointing you toward some meta-questions that I hope you will think 
about during the course of the summer institute. I call them to your 
attention because some wise people (including wise people from our 
own community, and even wise people from Stanford, California) have 
been pondering them as questions that might be moving toward 
scientific questions, to which we may hope to find scientific answers.
\begin{description}
    \item[$\rhd$]  Is this the best of all possible worlds? Dr. 
    Pangloss's assertion, though burdened with ironical baggage, 
    carries with it the daring suggestion that other worlds are 
    thinkable~\cite{candide}. According to an enduring dream that has probably 
    infected all of us from time to time, the theory of the world 
    might prove to be so restrictive that things have to turn out the 
    way we observe them. Is this really the way the world works, or 
    not? Are the elements of our standard model---the quarks and 
    leptons and gauge groups and coupling constants---inevitable, at 
    least in a probabilistic sense, or did it just happen this 
    way?\footnote{The paradigm of the ``string theory landscape'' offers 
    a very particular take on this 
    question~\cite{Susskind:2003kw,nimassi}. The string-theory landscape and
anthropic cosmological arguments~\cite{linde} seem to me to fall in the
tradition of Charles Sanders Peirce's ``Design and Chance,''
in \textit{The Essential Peirce,}
vol.  1 (1867 -- 1893),  ed.  Nathan Houser and Christian Kloesel
(University of Indiana Press, Bloomington \& Indianapolis, 1992), p. 215.
For a brief description of Peirce's evolutionary
cosmology, see Louis Menand, \textit{The Metaphysical Club: A Story of
Ideas in America} (Farrar Straus Giroux, New York, 2001), 
pp.~275--280. For a pithy critique of the anthropic approach, see the 
short essay by Paul Steinhardt at 
\url{http://www.edge.org/q2005/q05_print.html\#steinhardt}.}

    \item[$\rhd$] Is Nature simple or complex?  And if we take the
    sophisticate's view that it is both, which aspects will have
    beautiful ``simple'' explanations and which explanations will
    remain complicated?
    
    \item[$\rhd$]  Are Nature's Laws the same at all times and
    places? Yes, of course they are, to good approximation \textit{in our 
    experience.} Otherwise science would have had to confront a 
    universe that is in some manner capricious. But \textit{all} times and \textit{all} 
    places is a very strong conclusion, for which we cannot have 
    decisive evidence. Many people have been thinking about multiple 
    universes in which there may be different incarnations of the 
    basic structures.\footnote{For one provocative definition of 
    universes, see Ref.~\cite{Bjorken:2004an}.}

    \item[$\rhd$] Can one theoretical structure account for
    ``everything,'' or should we be content with partial theories
    useful in different domains? Can we really expect\footnote{This issue has 
    been joined recently by Freeman Dyson~\cite{dyson} and Brian 
    Greene~\cite{greene}.} to have a 
    theory that applies from the lowest energies to the highest, from 
    the smallest distances to the greatest?
\end{description}
All these questions are a bit wooly and may even be undecidable; they 
could generate a lot of blather and not lead to any telling insights. 
But we would be mistaken to pretend they are not there. So I urge 
you to spend a little of your time at the summer institute thinking 
about what constitutes a scientific explanation.

To work toward your own understanding of the Galilean relationship between 
small questions and sweeping insights, and to practice presenting the 
significance of your work to the wider world, please complete 
the following
\begin{center}
    \begin{boxit}
    \textit{Exercise.}  Explain in a paragraph or two how your current research
    project relates to Great Questions about Nature or is otherwise
    irresistibly fascinating.  Be prepared to present your answer to a
    science writer at a SSI social event.
    \end{boxit}
\end{center}

\section{ANTICIPATION}
\subsection{A Decade of Discovery Past\label{subsec:discovery}}
Before I move on to explore some themes that bind together the
questions that our organizers have given us (and some other topics), I
want to emphasize again that we stand on the threshold of a great
flowering of experimental particle physics and of dramatic progress in
theory---especially that part of theory that engages with experiment.

We particle physicists are impatient and ambitious people, and so we
tend to regard the decade just past as one of consolidation, as opposed
to stunning breakthroughs.  But an objective  look at the headlines of the past ten
years gives us a very impressive list of discoveries.  It is important
that we know this for ourselves, and that we convey our sense of
achievement and promise to others.\footnote{The citations that follow 
are to pertinent lectures at this school, rather than to the original 
literature.}

\begin{itemize}
    \addtolength{\itemsep}{-9pt}
    \item[$\rhd$] The electroweak theory has been elevated from a very 
    promising description to a \textit{law of nature.} It is quite 
    remarkable that in a  short time we have gone from a conjectured 
    electroweak theory to one that is established as a real quantum 
    field theory, tested as a quantum field theory at the level of 
    one per mille in many many observables~\cite{marciano}. This 
    achievement is truly the work of many hands; it has involved 
    experiments at the $Z^{0}$ pole, the study of $e^{+}e^{-}$, 
    $\bar{p}p$, and $\nu N$ interactions, and supremely precise 
    measurements such as the determination of 
    $(g-2)_{\mu}$~\cite{shagin}.
    
    \item[$\rhd$] Electroweak experiments have observed what we may 
    reasonably interpret as the influence of the Higgs boson in the 
    vacuum~\cite{marciano,snyder,deroeck}.
    
    \item[$\rhd$] Experiments using neutrinos generated by cosmic-ray 
    interactions in the atmosphere, by nuclear fusion in the Sun, and 
    by nuclear fission in reactors, have established neutrino flavor 
    oscillations: $\nu_{\mu} \to \nu_{\tau}$ and  $\nu_{e} \to 
    \nu_{\mu}/\nu_{\tau}$~\cite{gratta,casper,yokayama,waller,kayser}. 
    
    \item[$\rhd$] Aided by experiments on heavy quarks, studies of 
    $Z^{0}$,  investigations of high-energy $\bar{p}p$, $\nu N$, and $ep$ 
    collisions, and by developments in lattice field theory, we have 
    made remarkable strides in understanding quantum chromodynamics
    as the theory of the strong interactions.
    
    \item[$\rhd$] The top quark, a remarkable apparently elementary 
    fermion with the mass of an osmium atom, was discovered in 
    $\bar{p}p$ collisions~\cite{miller,clement}.
    
    \item[$\rhd$] Direct $\mathcal{CP}$ violation has been observed in $K \to \pi\pi$ decay. 
    
    \item[$\rhd$] Experiments at asymmetric-energy $e^{+}e^{-} \to 
    B\bar{B}$ factories have established that $B^{0}$-meson decays do 
    not respect $\mathcal{CP}$ invariance~\cite{lanceri}.
    
    \item[$\rhd$] The study of type-Ia supernovae and detailed thermal
    maps of the cosmic microwave background reveal that we live in an
    approximately flat universe dominated by dark matter and 
    energy~\cite{baltz,gaitskell,cabrera,rosenberg,refregier}.
    
    \item[$\rhd$] A ``three-neutrino'' experiment has detected the 
    interactions of tau neutrinos.
    
    \item[$\rhd$] Many experiments, mainly those at the highest-energy 
    colliders, indicate that quarks and leptons are structureless on 
    the \onetev.
\end{itemize}
We have learned an impressive amount in ten years, and I find quite 
striking the diversity of experimental and observational approaches 
that have brought us new knowledge, as well as the richness of the 
interplay between theory and experiment. 

Now I want to talk about five themes that weave together the great 
questions and small that we will be talking about during these two 
weeks.

\section{UNDERSTANDING THE EVERYDAY}
The first theme is one on which I am rather confident that we will 
make enormous progress over the next decade. That is the problem of 
understanding the everyday, the stuff of the world around us. It 
pertains to basic questions:  Why are there atoms?  Why 
is there chemistry?  Why are stable structures possible?  
And even, knowing the answers to those questions gives us an insight 
into What makes life possible?

Those are the general questions that we are seeking to answer when we 
look for the origin of electroweak symmetry breaking. I think that the 
best way to make the connection is to consider what the world would be 
like if there were no mechanism, like the Higgs mechanism, for
electroweak symmetry breaking. It's important to look at the problem 
in this way, because in the public presentations of the aspiration of 
particle physics we hear too often that the goal of the LHC or a 
linear collider is to check off the last missing particle of the 
standard model, this year's Holy Grail of particle physics, the Higgs 
boson. \textit{The truth is much less boring than that!} What we're 
trying to accomplish is much more exciting, and asking what the world 
would have been like without the Higgs mechanism is a way of getting 
at that excitement.

First, it's clear that quarks and leptons would remain massless,
because mass terms are not permitted in our left-handed world if the
electroweak symmetry remains manifest.\footnote{I assume for this
discussion that all the trappings of the Higgs mechanism, including
Yukawa couplings for the fermions, are absent.} We've done nothing to
QCD, so that would still confine the (massless) color-triplet quarks
into color-singlet hadrons, with very little change in the masses of
those stable structures.  In particular, the nucleon mass would be
essentially unchanged, but the proton would outweigh the neutron
because the down quark now does not outweigh the up quark, and that
change will have its own consequences.

An interesting, and slightly subtle point is that, even in the
absence of a Higgs mechanism, the electroweak symmetry is broken by
QCD, precisely by one of the emergent phenomena we have just discussed
in \S\ref{subsec:TOE}~\cite{Weinstein:1973gj}.  As we approach low
energy in QCD, confinement occurs and the chiral symmetry that treated
the massless left-handed and right-handed quarks as separate objects is
broken.  The resulting communication between the left-handed and
right-handed worlds engenders a breaking of the electroweak symmetry.
The trouble is that the scale of electroweak symmetry breaking is
measured by the pseudoscalar decay constant of the pion, so the amount
of mass acquired by the $W$ and $Z$ is set by $f_{\pi}$, not by what we
know to be the electroweak scale: it is off by a factor of 2500.  

But the fact is that the electroweak symmetry is broken, so the world
without a Higgs mechanism---but with strong-coupling QCD---is a world
in which the $\mathrm{SU}(2)_{\mathrm{L}}\otimes \mathrm{U}(1)_{Y}$
becomes $\mathrm{U}(1)_{\mathrm{em}}$.  Because the $W$ and $Z$ have
masses, the weak-isospin force, which we might have taken to be a
confining force in the absence of symmetry breaking, is not confining.
Beta decay is very rapid, because the gauge bosons are very light.  The
lightest nucleus is therefore one neutron; there is no hydrogen atom.
There's been some analysis of what would happen to big-bang
nucleosynthesis in this world; that work suggests that some light elements
such as helium would be
created~\cite{Agrawal:1998xa,Agrawal:1997gf,Hogan:1999wh,Yoo:2002vw}.\footnote{It
would be interesting to see this worked out in complete detail.} 
Because the electron is massless, the Bohr radius of the atom is 
infinite, so there is nothing we would recognize as an atom,  there 
is no chemistry as we know it, there are no stable composite 
structures like the solids and liquids we know.

I invite you to explore this scenario in even greater detail.  [To do
so is at least as challenging as trying to understand the world we do
live in.]  The point is to see how very different the world would be,
if it were not for the mechanism of electroweak symmetry breaking whose
inner workings we intend to explore and understand in the next decade.  What
we are really trying to get at, when we look for the source of
electroweak symmetry breaking, is why we don't live in a world so
different, why we live in the world we do.  I think that's a glorious
question. It's one of the deepest questions that human beings have ever
tried to engage, and \textit{you} will answer this question.

What could the answer be? As far as we can tell, because we have an 
\textit{effective field theory} description, the agent of 
electroweak symmetry breaking represents a novel fundamental 
interaction at an energy of a few hundred GeV. As we parametrize it 
in the standard electroweak theory, and we contrive the Higgs 
potential, it is not a gauge force but a completely new kind of 
interaction. We do not know what that force is.

What could it be? It could be the Higgs mechanism of the standard 
model, which is built in analogy to the Ginzburg--Landau description 
of superconductivity. Maybe it is a new gauge force. One very 
appealing possibility---at least until you get into the details---is 
that the solution to electroweak symmetry breaking will be like the 
solution to the model for electroweak symmetry breaking, the 
superconducting phase transition. The superconducting phase transition 
is first described by the Ginzburg--Landau phenomenology, but then in 
reality is explained by the Bardeen--Cooper--Schrieffer theory that 
comes from the gauge theory of Quantum Electrodynamics. Maybe, then, 
we will discover a mechanism for electroweak symmetry breaking almost 
as economical as the QCD mechanism we discussed above. One line that 
people have investigated again and again is the possibility that 
there are new constituents still to be discovered that interact by 
means of forces still to be discovered, and when we learn how to 
calculate the consequences of that theory we will find our analogue 
of the BCS theory. It could even be that there is some truly emergent 
description---at this level---of the electroweak phase transition, a residual force 
that arises from the strong dynamics among the weak gauge bosons. 
We know that if we take the mass of the Higgs boson to very large 
values, beyond a TeV in the Lagrangian of the electroweak theory, the 
scattering among gauge bosons becomes strong, in the sense that 
$\pi\pi$ scattering becomes strong on the GeV scale. Resonances form 
among pairs of gauge bosons, multiple production of gauge bosons 
becomes commonplace, and that resonant behavior could be what hides 
the electroweak symmetry. We'll also hear during these two weeks about 
the possibility that electroweak symmetry breaking is the echo of 
extra spacetime dimensions. We don't know, and we intend to find out 
during the next decade which path nature has taken.

One very important step toward understanding the new force is to find 
the Higgs boson and to learn its properties. I've said before in 
public, and I say again here, that the Higgs boson will be discovered 
whether it exists or not. That is a statement with a precise 
technical meaning. There will be (almost surely)  a spin-zero 
object that has effectively more or less the interactions of the 
standard-model Higgs boson, whether it is an elementary 
particle that we put into to the theory or something that emerges from 
the theory. Such an object is required to get good high-energy 
behavior of the theory. 

If something will be found, what is it? How many are there? Is its 
spin-parity what we expect ($J^{PC} = 0^{++}$) in the electroweak 
theory? Does it generate mass for the gauge bosons $W$ and $Z$ alone, 
or does it generate mass for the gauge bosons and the fermions? How 
does it interact with itself? 

There will be a party on the day the Higgs boson is discovered, but it will 
mark the beginning of a lot of work!

\section{THE MEANING OF IDENTITY}
The second theme has to do with the cast of characters, the basic
constituents of matter, the quarks and leptons.  It involves the
question, ``What makes a top quark a top quark, an electron an
electron, a neutrino a neutrino? What distinguishes these objects?'' 
Now, maybe this is a Kepler-style question that we shouldn't be 
asking, but it is a tantalizing question in any event.

What do I mean by this more precisely?  I mean, what sets the masses
and mixings of the quarks and leptons?  This has to do with the famous
CKM matrix of quark mixings, which our colleagues here and elsewhere
are measuring so assiduously.  These elements arise, in the standard
model, in the course of electroweak symmetry breaking with values set
by those famous arbitrary Yukawa couplings, whose values we don't know
except by experiment.  What is $\mathcal{CP}$ violation really trying to
tell us?  One of the things I am most confused about is what discrete
symmetries mean, when they are exact and when they are broken. Are 
parity violation and $\mathcal{CP}$ violation intrinsic defects---or 
essential features---of the laws of nature, or do they represent 
spontaneously broken symmetries?

Neutrino oscillations---flavor-changing transitions, more
generally---give us a new look at the meaning of identity, because they, too, have
to do with fermion masses and identities.  Neutrino masses can be
generated in the old ways, through Yukawa couplings, and in new ways as
well~\cite{Mohapatra:2004ht}, so they may give us a new take on the problem, and add richness
to it.  We often hear that neutrino mass is evidence for physics beyond
the standard model.

I'm here to tell you that \textit{all fermion masses, starting with the
electron mass, are evidence for physics beyond the standard model.} The
reason in this: while, in the electroweak theory a little box pops up
and says, ``Write the electron mass here,'' nothing in the electroweak
theory---either now or at any time in the future---is going to tell us
how to calculate that number.  It's not that the calculation is
technically challenging, it is that the electroweak theory has nothing
to say about fermion mass.  All of these masses are profoundly
mysterious.  Neutrino masses could present an additional mystery,
because neutrinos can be their own antiparticle, which means there are
other ways of generating neutrino masses.  There is a real enigma here,
one that we need to get our minds around.

Maybe we haven't figured out what the pattern is because there is 
more to see in the pattern~\cite{tata,wang,desch,okada}. Perhaps it will only become apparent when 
we take into account the masses of superpartners or other kinds of 
matter. It's worth remembering that when Mendele'ev made his periodic 
table, he constructed it out of the chemical elements that had been 
discovered by chemists. The chemicals discovered by chemists are the 
chemicals that have chemistry; and so Mendele'ev didn't know about 
helium, neon, argon, krypton, xenon. If you had tried to see the 
pattern, you would have made real progress filling in the missing 
elements, but without the noble gases that we now think of as the last 
column, you wouldn't have had the clues necessary to build up, in a 
systematic way, the properties of the elements, or to guess what lies 
behind the periodic table. Perhaps we need to see something more---an 
analogue of the noble gases---before we can understand what lies 
behind the pattern.

I'm less confident that in ten years we will get to the bottom of this
theme, because I really think that we are at the stage of developing
for ourselves what this question is.  We know very well what are the
measurements we'd like to make in $B$ physics, charm and strange
physics, and neutrino physics---which elements of the mixing matrices
we would like to fill in and which relationships we would like to test.
But I don't think we've done a satisfactory job yet of constructing
what the big question is, and what the properties of the fermions are
trying to tell us.  I think it is very important that we try to think
of the quarks and leptons together, to see what additional insights a
common analysis might bring, and to try to understand what the question
really is here.

Among the extensions to the standard model that might give us clues 
into the larger pattern there is, of course, supersymmetry. In common 
with many extensions to the standard model, supersymmetry brings us 
dark matter candidates~\cite{matchev}. Supersymmetry is very highly developed. It 
has a number of very important consequences if it is true.\footnote{By which 
I mean, if it is true and relevant on the 1-TeV scale. Supersymmetry 
might be true and shape physics on the Planck scale but have nothing 
directly to do with these issues.}
First, if the top quark is heavy and a few other things happen in the 
right way, then supersymmetry predicts the condensation that gives 
rise to the hiding of  electroweak symmetry. It can generate, by the running 
of masses, the shape of the Higgs potential. It predicts a light Higgs 
mass, less than some number in the neighborhood of 130, 140, 
150~GeV. That's consistent with the current indications from 
precision electroweak measurements. It predicts cosmological cold dark 
matter, which seems to be a good thing to have. It might lead to an 
understanding of the excess~\cite{coppi,streitmatter} of matter over 
antimater in the universe~\cite{wagner,Trodden:2004mj}. And, in a unified 
theory, it explains the (relative) values of the standard-model 
coupling constants.  To see that, we have to move on to the 
next theme.

\section{THE UNITY OF QUARKS AND LEPTONS}
The quarks have strong interactions, as you all know, and the leptons
don't.  Could we have a world made only of quarks, or only of leptons?
There are many strong reasons for believing that quarks and leptons
must have something to do with each other, despite their different
behavior under the strong interactions.  What do they have in common?
They are all spin-$\cfrac{1}{2}$ particles, structureless at the
current limits of resolution.  The six quarks match the six leptons.
What motivates us to think of a world in which the quarks and leptons
are not just unrelated sets that match by chance, but have a deep
connection?  The simplest way to express it, I think, is to go back to
a puzzle of very long standing, why atoms are so very nearly neutral.
This is one of the best measured numbers close to zero in all of
experimental science: atoms are neutral to one part in $10^{22}$.

If there is no connection between quarks and leptons, since quarks make
up the proton, then the balance of the proton and electron charge is
just a remarkable coincidence.  It seems impossible for any thinking
person to be satisfied with coincidence as an explanation.  Some
principle must relate the charges of the quarks and the leptons.  What
is it?  A fancier way of saying it, and more or less equivalent, is
that for the electroweak theory to make sense up to arbitrarily high
energies, the symmetries on which it is based must survive quantum
corrections.  The way we say that is that the theory must be free of
anomalies---quantum corrections that break the gauge symmetry on which
the theory is based.  In our left-handed world, that is only possible
if weak-isospin pairs of color-triplet quarks accompany weak-isospin
pairs of color-singlet leptons.  For these reasons, it is nearly
irresistible to consider a unified theory that puts quarks and leptons
into a single extended family.

Once you've done that, it's a natural implication that protons should
decay.  Although it's a natural implication, it may not be unavoidable,
because we don't know which quarks go with which leptons.  If you look
at the tables chiseled in marble out in the hallway to celebrate the
Nobel Prize of 1976, you will see that the up and down quarks go with
the electron and its neutrino.  We have no experimental basis for that
arrangement, it just reflects the order in which we met the particles.
For all we know, the first generation of quarks goes with the third
generation of neutrinos.  Supersymmetry is interesting in this context
because it sets an experimental target that's not so far away---an
order of magnitude or two away: Perhaps that target provides enough
stimulus---if we can think of how to build a massive, low-background
apparatus at finite cost---to go the next order of magnitude or two in
sensitivity, perhaps to find evidence for proton decay, which would be
the definitive proof of the connection between quarks and leptons.

Coupling constants unify in the unified theory.  At some high scale,
whose value we might discover in some future theory, all the couplings
have a certain value.  The differing values we see at low energy for
the $\mathrm{U}(1)$ associated with weak hypercharge, the
$\mathrm{SU}(2)$ associated with weak isospin, and the $\mathrm{SU}(3)$
associated with color come about because of the different evolution
given by the different gauge groups and the spectrum of particles
between up there and down here.  In this sense we can explain why the
strong interactions are strong on a certain scale.

One way of thinking about the masses of the quarks and leptons is to 
imagine that the pattern just looks weird to us because we are 
examining the fermion masses at low energies. Masses run with 
momentum scale in a way analogous to the running of coupling 
constants. So possibly, if we look at very high energies, we will see 
a rational pattern that relates one mass to another through 
Clebsch--Gordan coefficients or some other symmetry factors. There are 
examples of this. One of the nice fantasy studies for the linear 
collider is measuring masses of  superpartners 
well enough at low energies to have the courage to extrapolate them 
over fourteen or fifteen orders of magnitude in energy, to see how 
they come together~\cite{Allanach:2004ud}.

\section{GRAVITY REJOINS PARTICLE PHYSICS REJOINS GRAVITY REJOINS \ldots}
We particle physicists have neglected gravity all these years, and for
good reason.  If we calculate a representative process,  kaon
decay into a pion plus a graviton for example, it's easy to estimate that the
emission of a graviton is suppressed by  $M_{K}/M_{\mathrm{Planck}}$.
  The Planck mass ($M_{\mathrm{Planck}} \equiv (\hbar 
  c/G_{\mathrm{Newton}})^{1/2}\approx 1.22
\times 10^{19}\gev$) is a big number because Newton's constant is small in the
appropriate units.  A dimensional estimate for the branching fraction
is $B(K \to \pi G) \approx ({M_{K}}/{M_{\mathrm{Planck}}})^{2} \approx
10^{-38}$. It will be a long time before the single-event sensitivity 
of any kaon experiment reaches this level! And that's why we have 
been able to safely neglect gravity most of the time.

All of us have great respect for the theory of gravity, because it was
given to us by Einstein and Newton and the gods, whereas we know the
people who made the electroweak theory, and so it's natural to think
that gravity must be true.  But from the experimental point of view, we
know very little about gravity at short
distances~\cite{Arkani-Hamed:1998rs}.  Down to a few tenths of a
millimeter, elegant experiments~\cite{Adelberger:2003zx,smullin} using
torsion oscillators and microcantilevers exclude a deviation from
Newton's inverse-square law with strength comparable to gravity's.  The
techniques and the bounds are very impressive!  But at shorter
distances, the constraints deteriorate rapidly, so nothing prevents us
from considering changes to gravity even on a small but macroscopic
scale.  Even after this new generation of experiments, we have only
tested our understanding of gravity---through the inverse-square
law---up to energies of 10~meV (yes, \textit{milli}-electron volts),
some fourteen orders of magnitude below the energies at which we have
studied QCD and the electroweak theory.  That doesn't mean that a
deviation from the inverse-square law is just around the corner, but
experiment plainly leaves an opening for gravitational surprises. 
Indeed, it is an open possibility that at \textit{larger} distances than we have observed 
astronomically gravity might deviate from the inverse-square law.
There is a huge field over which gravity might be different from
Newton's law, and we wouldn't have discovered it yet.

Now, in spite of the fact that we have had good reason to neglect
gravity in our daily calculations of Feynman diagrams, we have also
been keenly aware that gravity is not always negligible.  In more or
less any interacting field theory, and certainly in one like the
electroweak theory, where the Higgs field has a nonzero value that
fills all of space, all of space has some energy density.  In the
electroweak theory, that energy density turns out to be really large.
If you calculate it, you find that the contribution of the Higgs
field's vacuum expectation value to the energy density of the universe
is $\varrho_{H} \equiv {M_{H}^{2}v^{2}}/{8}$, where $M_{H}$ is the
Higgs-boson mass and $v \approx 246\gev$ is the scale of electroweak
symmetry breaking.  A vacuum energy density corresponds to a
cosmological constant $\Lambda = { ({8\pi
G_{\mathsf{Newton}}}/{c^{4}}}){\varrho_{\mathsf{vac}}}$ in Einstein's
equations.  We've known for a very long time that there is not much of
a cosmological constant, that the vacuum energy has to less than about
$\varrho_{\mathsf{vac}} \lesssim 10^{-46}\gev^{4}$, a very little
number.  It corresponds to $\approx 10\mev/\ell$ or $10^{-29}\hbox{
g}\cm^{-3}$.  Even in the blackest heart, there is not much dark
energy!

But if we use the current lower limit on the 
Higgs-boson mass,  $M_{H} \gtrsim 114\gev $, to estimate the vacuum 
energy in the electroweak theory, we find $\varrho_{H} \gtrsim  10^{8}\gev^{4}$ .
That is wrong by no less than fifty-four orders of magnitude! This mismatch has 
been known for about three decades. That long ago, Tini Veltman was 
concerned that something fundamental was missing from our 
conception of the electroweak theory. For many of us, the vacuum 
energy problem has been a chronic dull headache for all this time.

This raises an interesting point about how science is done, and how 
science progresses. We could, all of us, have said, ``The electroweak 
theory is wrong, let's put it aside.'' Think of all that we 
wouldn't know, if we had followed that course. We can't forget about 
deep problems like the vacuum energy conflict, but we have to have the 
sense to put them aside, to defer consideration until the right 
moment. In the simplest terms, the question is, ``Why is empty space so 
nearly massless?'' That is a puzzle that has been with us repeatedly 
in the history of physics, and it is one that is particularly pointed 
now. Maybe now should be the time that we return to the vacuum 
energy problem.

Over the last few years, we have a new wrinkle to the vacuum energy
puzzle, the evidence---within a certain framework of analysis---for
 a nonzero cosmological constant, respecting the bounds cited a
moment ago.  That discovery recasts the problem in two important ways.
First, instead of looking for a principle that would forbid a
cosmological constant, perhaps a symmetry principle that would set it
exactly to zero, now we have to explain a tiny cosmological constant!
Whether we do that in two steps or one step remains to be seen.
Second, from the point of view of the dialogue among observation and
experiment and theory, now it looks as if we have access to some new
stuff whose properties we can measure.  Maybe that will give us the
clue that we need to solve this old problem.

We now come to the question of how we separate the electroweak scale 
from higher scales~\cite{lykken}. This is a realm in which we haven't neglected 
gravity all along, because we have wanted to think of the electroweak 
theory as a truly useful  effective theory, 
and we have known that we live in a world in which the electroweak 
scale isn't the only scale. We have taken note of the Planck scale, 
and there may be a unification scale for strong, weak, and electromagnetic interactions;
for all we know, there are intermediate scales, where flavor 
properties are determined and masses are set. 

We know that the Higgs-boson mass must be less than a TeV, but the 
scalar mass
communicates quantum-mechanically with the other scales that may range
all the way up to $10^{19}\gev$.  How do we keep the Higgs-boson mass
from being polluted by the higher scales?  That's the essence of the
hierarchy problem.  We've dealt with this, for twenty-five years or so,
by extending the standard model.  Maybe the Higgs boson is a composite
particle, maybe we have broken supersymmetry that tempers the quadratic
divergences in the running of the Higgs-boson mass, maybe \ldots .
Now, because of the observation that we haven't tested gravity up to
very high energies, it has become fashionable to turn the question
around and ask why the Planck scale is so much bigger than the
electroweak scale, rather than why the electroweak scale is so low.  In
other words, why is gravity so weak?

\section{A NEW CONCEPTION OF SPACETIME}
That line of investigation has given rise to new thinking, part of it
connected with a new conception of spacetime.  What is in play here,
again, is a question so old that, for a long time, we had forgotten
that it was a question: Is spacetime really three-plus-one dimensional?
What is our evidence for that?  How well do we know that there are not
other, extra, dimensions?  What must be the character of those extra
dimensions, and the character of our ability to investigate them, for
them to have escaped our notice?

Could extra dimensions be present?  What is their size?  What is their
shape?  What influence do they exert on our world?  (Because if they
have no effect, it almost doesn't matter that they exist.)  Are the
extra dimensions where fermion masses are set, or electroweak symmetry
is broken, or what?  How can we map them?  How can we attack the
question of extra dimensions experimentally?

I will give you just two examples of new ways of thinking that are 
stimulated by the notion that additional dimensions have eluded 
detection. These are both probably wrong, and that hardly 
matters, because they are mind-expanding. 

Perhaps, in contrast to the strong and electroweak gauge forces,
gravity can propagate in the extra dimensions---in all dimensions,
because it is universal.  When we inspect the world on small enough
scales, we will see gravity leaking into the extra dimensions.  Then by
Gauss's law, the gravitational force will not be an inverse-square law,
but will be proportional to $1/r^{2+n}$, where $n$ is the number of
extra dimensions.  That would mean that, as we extrapolate to smaller
distances, or higher energies, gravity will not follow the Newtonian
form forever, as we conventionally suppose. Below a certain
distance scale, it will start evolving more rapidly; its strength will
grow faster.  Therefore it might join the other forces at a much lower
energy than the Planck scale we have traditionally assumed.  That could change our
perception of the hierarchy problem entirely.  That's a way we hadn't
thought about the problem before.  It has stimulated a lot of research
into how we might detect extra dimensions~\cite{Rizzo:2004kr,Landsberg:2004mj}.

Perhaps extra dimensions offer a new way to try to understand fermion
masses~\cite{Arkani-Hamed:1999dc}.  One of the great
challenges---beyond the fact that we don't have a clue how to calculate
fermion masses---is that the fermion masses have such wildly different
values.  In units of $v/\sqrt{2}$, the mass of the top quark is $1$,
the mass of the electron is a few $\times 10^{-6}$, and so on.  How can
a reasonable theory generate such big differences?  Suppose, for
simplicity, that spacetime has one additional dimension.  In that extra
dimension,  wave packets correspond to left-handed and
right-handed fermions.  For reasons to be supplied by a future theory,
each wave packet rides on a different rail (is centered on a different
value of the new coordinate, $x_{\mathrm{new}}$).  It is the overlap
between a left-handed wave packet, a right-handed wave packet, and the
Higgs field---assumed to vary little with $x_{\mathrm{new}}$---that
sets the masses of the fermions.  If the wave packets are Gaussian (how
else could they be?)  then they need only be offset by a little in
order for the overlap integral to change by a lot.  I don't know
whether this story can possibly be right, but it is very different from
any other story we have told ourselves about fermion masses.  For that
reason, I think it is an important opening.

Other extra-dimensional delights may present themselves, provided that
gravity is intrinsically strong but spread out into many dimensions.
Tiny black holes might be formed in high-energy
collisions~\cite{giddings}.  We might have the possibility of detecting
the exchange or emission of gravitons---not as individual gravitons,
but as towers of them~\cite{Hewett:2002hv}.\footnote{In the colloquy
cited in \S\ref{subsec:metaq}, Freeman Dyson asserts that we don't need
a quantum theory of gravity because single graviton emission can never
be detected.  We would say that he is mistaken, but the dialogue
reveals an interesting contrast of styles and world-views.} At all
events, gravity is here to stay in particle physics.  It's been present
for years as a headache, in the form of the hierarchy problem and in
the challenge of the vacuum energy problem.  Now it is perhaps
presenting itself as an opportunity!

\section{THE DOUBLE SIMPLEX}
As I intimated in \S\ref{subsec:natq}, I have been concerned for some
time with the prevailing narrow view of the goals of our science.  It
is troubling, to be sure, when we read in the popular press that the
sole object of our endeavors is to find---to check off, if you
will---the Higgs boson, the holy grail (at least for this month) of
particle physics.\footnote{I say this as someone whose obsession with
electroweak symmetry breaking is no secret!} What is more troubling to
me, the shorthand of the Higgs search narrows the discourse within our
own community.  In response, I have begun to evolve a visual
metaphor---the double simplex---for what we know, for what we hope
might be true, and for the open questions raised by our current
understanding.  While I have a deep respect for the refiner's fire that
is mathematics, I believe that we should be able to explain the essence
of our ideas in languages other than equations.  I interpolated a brief
animated overview~\cite{EnvPIQT} of the double simplex\footnote{Any
resemblance to Kepler's \textit{stella octangula} is purely
coincidental.} at this point in my lecture. For a preliminary 
exposition in a pedagogical setting, see Ref.~\cite{Quigg:2004is}. A 
more complete explanation of the aims of particle physics through the 
metaphor of the double simplex is in preparation.

\section{ANTICIPATION}

\subsection{A decade of discovery ahead}
I spoke at the beginning of the hour about the decade of discovery just
achieved.  I believe that the decade ahead will be a real golden age of
exploration and discovery.

\begin{itemize}
    \addtolength{\itemsep}{-9pt}
    \item[$\rhd$] We will make a thorough exploration of the 1-TeV 
    energy scale; search for, find, and study the Higgs boson or its 
    equivalent; and probe the mechanism that hides electroweak 
    symmetry.  Decisive progress will come from our 
    (anti)proton-proton colliders, notably the Large Hadron Collider 
    at CERN, but we envisage a TeV-scale electron-positron linear 
    collider to give us a second look, through a different 
    lens.\footnote{It is wrong to say, as well-meaning people 
    sometimes do, that the LHC is a blunt instrument and the LC a 
    scalpel. A more apt analogy is to the suite of telescopes---radio, 
    infrared, optical, ultraviolet, X-ray, etc.---that enrich 
    astronomical observations. Each instrument is made more capable by 
    the dialogue with its companions.}

    \item[$\rhd$] We will continue to challenge the standard model's 
    attribution of $\mathcal{CP}$ violation to a phase in the quark 
    mixing matrix, in experiments that examine $B$ decays and rare 
    decays---or mixing--of strange and charmed particles. Fixed-target 
    experiments, as well as $e^{+}e^{-}$ and $p^{\pm}p$ colliders, 
    will contribute.
    
    \item[$\rhd$] New accelerator-generated neutrino beams, together 
    with reactor experiments and the continued study of neutrinos from 
    natural sources, will consolidate our understanding of neutrino 
    mixing. Double-beta-decay searches may confirm the Majorana 
    nature of neutrinos. And do not dismiss the possibility that three 
    neutrinos will not suffice to explain all observations!
    
    \item[$\rhd$] The top quark will become an important window into 
    the nature of electroweak symmetry breaking, rather than a mere 
    object of experimental desire. Single-top production and the top 
    quark's coupling to the Higgs sector will be informative. Hadron 
    colliders will lead the way, with the LC opening up additional 
    detailed studies.
    
    \item[$\rhd$] The study of new phases of matter and renewed 
    attention to hadronic physics will deepen our appreciation for the 
    richness of QCD, and might even bring new ideas to the realm of 
    electroweak symmetry breaking.\footnote{It bears repeating that 
    we owe most of our ideas about electroweak symmetry breaking to 
    the superconducting phase transition.} Heavy-ion collisions have 
    a special role to play here, but $ep$ collisions, fixed-target 
    experiments, and $p^{\pm}p$ and $e^{+}e^{-}$ colliders all are 
    contributors.
    
    \item[$\rhd$] Planned discoveries and programmatic surveys have 
    their (important!) place, but exploration breaks the mold of 
    established ideas and can recast our list of urgent 
    questions overnight. The LHC, not to mention a whole range of 
    experiments down to tabletop scale, will make the coming decade one of 
    the great voyages into the unknown. Among the objectives we have 
    already prepared in great theoretical detail are extra 
    dimensions, new strong dynamics, supersymmetry, and new forces 
    and constituents. Any one of these would give us a new continent 
    to explore.
    
    \item[$\rhd$] Proton decay remains the most promising path to 
    establish the existence of extended families that contain both 
    quarks and leptons. Vast new underground detectors will be 
    required to push the sensitivity frontier.
    
    \item[$\rhd$] We will learn much more about the composition of 
    the universe, perhaps establishing the nature of some of the dark 
    matter. Observations of type Ia supernovae, the cosmic microwave 
    background, and the large-scale structure of the universe will 
    extend our knowledge of the fossil record. Underground searches 
    may give evidence of relic dark matter. Collider experiments will 
    establish the character of dark-matter candidates and will make 
    possible a more enlightened reading of the fossil record.
\end{itemize}

These few items constitute a staggeringly rich prospectus for search 
and discovery and for enhanced understanding. Exploiting all these 
opportunities will require many different instruments, as well as the 
toil and wit of many physicists. Fred
Gilman~\cite{gilman} will offer a roadmap to the future at the end of
the school, but it is plain that one of our great challenges is to
think clearly about the diversity of our experimental initiatives, and
about scale diversity of those initiatives.  It is relatively easy to
write the major headlines of the program we would like to see.  But how
do we create the institutions that year after year make important
measurements?  How do we create the next set of Greatest Puzzles?
That, it seems to me, is a very significant issue for people who will
be part of our field over the next thirty years.

I leave you with a list of advances that I believe can happen over the 
next decade or so. I put up my list for the same reason, I think, 
that the organizers of the school gave you their list---because then 
you can object to it, and make your own! We will \ldots
\begin{quote}
Understand electroweak symmetry breaking,
		Observe the Higgs boson,
		Measure neutrino masses and mixings,
		Establish Majorana neutrinos through the observation of neutrinoless double-beta decay,
		Thoroughly explore $\mathcal{CP}$ violation in $B$ decays,
		Exploit rare decays ($K$, $D$, \ldots),
		Observe the neutron's permanent electric dipole meoment, and pursue the electron's
		electric dipole moment,
		Use top as a tool,
		Observe new phases of matter,
		Understand hadron structure quantitatively,
		Uncover the full implications of QCD,
		Observe proton decay,
		Understand the baryon excess of the universe,
		Catalogue the matter and energy of the universe,
		Measure the equation of state of the dark energy,
		Search for new macroscopic forces,
		Determine the gauge symmetry that unifies the strong, weak, and electromagnetic interactions,
	Detect neutrinos from the universe,
	Learn how to quantize gravity,
		Learn why empty space is nearly weightless,
		Test the inflation hypothesis,
		Understand discrete symmetry violation,
		Resolve the hierarchy problem,
		Discover new gauge forces,
		Directly detect dark-matter particles,
		Explore extra spatial dimensions,
		Understand the origin of the large-scale structure of the universe,
		Observe gravitational radiation,
		Solve the strong $\mathcal{CP}$ problem,
		Learn whether supersymmetry operates on the TeV scale,
		Seek TeV-scale dynamical symmetry breaking,
		Search for new strong dynamics,
		Explain the highest-energy cosmic rays,
		Formulate the problem of identity, \ldots
\end{quote}
\noindent		
\ldots and learn the right questions to ask!

\section{NATURE'S NEGLECTED PUZZLES}
I've given you my view of how our 
puzzles and opportunities and clues fit together, of how we might 
think about our field and  evolution.  The organizers have given you 
their picture, with  ten themes for  ten days of our school. To 
encourage lively participation and debate, I issued \ldots
\begin{center}
    \begin{boxit}
    \textit{The NNP Challenge:}  Propose a question not on the SSI2004 list, and
explain briefly why it belongs in the pantheon of Nature's Greatest
Puzzles~\cite{shawne}.
    \end{boxit}
\end{center}
The contest was open to any student at the 
SLAC Summer Institute---anybody  willing to propose a new question to 
 be judged by our international panel of experts. 

I  presented the reward for the Best Eleventh Question on Wednesday,
August 11 to SISSA/SLAC graduate student Yasaman
Farzan, for her question about the validity of Poincar\'{e}  
invariance:
\begin{center}
    \begin{boxit}
	\textit{To what extent is  Poincar\'{e} symmetry exact?}
    Looking back on the history of science, discovering that different
    symmetries are not exact has ushered in a new era.  Poincar\'{e} symmetry
    is particularly interesting because it is currently considered the most
    sacred geometry.  Moreover, its evolution to the form we learn about
    today has marked great revolution in physics, in the past.
    \end{boxit}
\end{center}
Yasaman's trophy, a bottle of California's finest sparkling
wine,\footnote{By my decree as organizer of the competition, 1996 Iron
Horse Brut LD.} bears the autographs of Nobel Laureates Martin Perl and
Burton Richter; SLAC notables Jonathan Dorfan, Persis Drell, Sid Drell,
and Vera L\"{u}th; High Energy Physics Advisory Panel Chair Fred
Gilman; SLAC Summer Institute organizers JoAnne Hewett, John Jaros,
Tune Kamae, and Charles Prescott; and my own.  Even more precious was the opportunity---need we say,
obligation--to present and defend the best eleventh question in an
eleven-minute talk at that day's afternoon Discussion Session.  Padova
student Marco Zanetti and Colorado State/UCSD student Thomas Topel
received special commendations for their questions on the nature of
time and the mechanism that breaks the strong--electroweak symmetry.
Their prizes are copies of Peter Galison's recent book,
\textit{Einstein's Clocks, Poincare's Maps: Empires of Time.}
Thanks and congratulations to all~\cite{NNP} who entered the Challenge!

\begin{acknowledgments}
Fermilab is operated by Universities Research Association Inc.\ under
Contract No.\ DE-AC02-76CH03000 with the U.S.\ Department of Energy.
I gratefully acknowledge the warm hospitality of CERN--TH, where I 
prepared the final form of these notes. I thank Tom Appelquist, 
Andreas Kronfeld, and Marvin Weinstein for insightful comments on 
emergent phenomena. My enthusiastic thanks go to the organizers and 
participants in the XXXII SLAC Summer Institute for a very enjoyable 
and educational fortnight.
\end{acknowledgments}


\end{document}